\documentclass[twocolumn,superscriptaddress,amsmath,amssymb,aps,prd]{revtex4-2}

\usepackage{graphicx}
\usepackage{dcolumn}
\usepackage{pifont}
\usepackage{enumitem} 
\usepackage{graphicx}
\usepackage{dcolumn}
\usepackage{bm}
\usepackage{mathtools} 
\usepackage{amsfonts}
\usepackage{booktabs}
\usepackage{siunitx}
\usepackage{verbatim}
\usepackage[normalem]{ulem} 
\usepackage[colorlinks=true,citecolor=blue,urlcolor=blue]{hyperref}
\usepackage[usenames,dvipsnames]{xcolor}
\usepackage{enumitem}

\newcommand{\prlsection}[1]{\paragraph*{#1 \textendash{}}}

\newcommand{\Austin}{\affiliation{Center for Gravitational Physics, University of Texas at Austin, Austin, Texas 78712, USA}}

\newcommand{\Cornell}{\affiliation{Cornell Center for Astrophysics and Planetary Science, Cornell University, Ithaca, New York 14853, USA}}

\newcommand{\Caltech}{\affiliation{Theoretical Astrophysics, Walter Burke Institute for Theoretical Physics,
California Institute of Technology, Pasadena, California 91125, USA}}

\begin{document}

\title{Success of the small mass ratio approximation during the final orbits of binary black hole simulations}

\author{Sergi Navarro Albalat}
\Austin
\author{Aaron Zimmerman}
\Austin
\author{Matthew Giesler}
\Cornell
\author{Mark A. Scheel}
\Caltech

\date{\today}

\begin{abstract}
Recent studies have shown the surprising effectiveness of the small mass-ratio approximation (SMR) in modeling the relativistic two-body problem even at comparable masses.
Up to now this effectiveness has been demonstrated only during inspiral, before the binary transitions into plunge and merger. 
Here we examine the binding energy of nonspinning binary black hole simulations with mass ratios from 20:1 to equal mass.
We show for the first time that the binaries undergo a transition to plunge as predicted by analytic theory, and estimate the size of the transition region, which is $\sim 10$ gravitational wave cycles for equal mass binaries.
By including transition, the SMR expansion of the binding energy is accurate until the last cycle of gravitational wave emission.
This is true even for comparable mass binaries such as those observed by current gravitational wave detectors, where the transition often makes up much of the observed signal.
Our work provides further evidence that the SMR approximation can be directly applied to current gravitational wave observations.
\end{abstract}

\maketitle

\prlsection{Introduction}
The direct detection of gravitational waves (GWs)~\cite{LIGOScientific:2016aoc,LIGOScientific:2018mvr,LIGOScientific:2020ibl,LIGOScientific:2021usb,LIGOScientific:2021djp,Nitz:2018imz,Nitz:2020oeq,Nitz:2021uxj,Nitz:2021zwj,Zackay:2019tzo,Venumadhav:2019tad,Venumadhav:2019lyq,Zackay:2019btq,Olsen:2022pin} has provided a new view of the universe, revealing populations of binaries composed of black holes and neutron stars~\cite{LIGOScientific:2018jsj,LIGOScientific:2020kqk,LIGOScientific:2021psn} and enabling tests of relativity in the dynamical and strong-field regime, e.g.~\cite{LIGOScientific:2016lio,Yunes:2016jcc,LIGOScientific:2016dsl,LIGOScientific:2018dkp,LIGOScientific:2019fpa,LIGOScientific:2020tif,LIGOScientific:2021sio}.
Accurate modeling of the relativistic two-body problem is crucial for carrying out sensitive GW searches and measuring the parameters of detected binaries.
For example effective one body models (e.g.~\cite{Buonanno:1998gg,Buonanno:2000ef,Taracchini:2013rva,Antonelli:2019fmq,Ossokine:2020kjp,Liu:2021pkr,Nagar:2021gss,Ramos-Buades:2021adz,Albertini:2021tbt}) and phenomenological models (e.g.~\cite{PhysRevD.104.124027,Ajith:2009bn,
Hannam:2013oca,Khan:2018fmp,Khan:2019kot,Pratten:2020fqn,Pratten:2020ceb})
predict the GWs emitted during the binary inspiral, merger, and post-merger phases.

The construction of such models requires input from a number of methods, such as post-Newtonian (PN), gravitational self-force (GSF) and numerical relativity (NR) techniques, each with different limitations. 
For example the PN approximation is valid at large separations and slow velocities.
It worsens late in the inspiral, and can be a poor approximation for systems with small mass ratios (SMRs) $q^{-1}$, where $q = m_1/m_2 \geq 1 $, which spend many cycles at close separations.
In contrast, for SMR binaries, GSF methods based on an expansion of the metric about a black hole background are more accurate.
GSF is well-suited to describe extreme-mass-ratio inspirals (EMRIs) with $q^{-1} \sim 10^{-4}$--$10^{-6}$,
which are promising targets for the LISA mission~\cite{Audley:2017drz}. It has recently been pushed to second-order (2GSF) for nonspinning binaries~\cite{Pound:2019lzj,Miller:2020bft,Warburton:2021kwk,Wardell:2021fyy}. Meanwhile, numerical relativity (NR) provides two-body solutions exact up to numerical errors, 
and can be used to assess the validity of the PN and SMR approximations in the regime of comparable masses and small separations accessible to NR~\cite{LeTiec:2014oez}. 
NR is also used to calibrate 
full inspiral-merger-postmerger 
models, or to build surrogates which can interpolate waveform predictions between simulations, e.g.~\cite{Field:2013cfa,Blackman:2015pia,Blackman:2017pcm,Varma:2019csw,Rifat:2019ltp,Islam:2022laz,Yoo:2022erv}. 

Surprisingly, comparisons between NR and SMR approximations of binary black hole systems have shown that the
latter is effective at describing even comparable mass ratio systems,
e.g.~\cite{LeTiec:2011bk,LeTiec:2011dp,LeTiec:2013uey,LeTiec:2017ebm,vandeMeent:2020xgc,Warburton:2021kwk,Wardell:2021fyy,Albalat:2022lfz}.
These include the binding energy, GW fluxes and GW phasing, and is achieved by re-expanding the SMR series in the symmetric mass ratio $\nu\equiv m_1m_2/(m_1+m_2)^2$ rather than $q^{-1}$~\cite{LeTiec:2013uey,LeTiec:2017ebm,Wardell:2021fyy}.
This promising result indicates that GSF methods could model 
GWs from binaries detectable by
current~\cite{LIGOScientific:2014pky,VIRGO:2014yos,KAGRA:2020tym} and future ground-based detectors~\cite{Amaro-Seoane:2018gbb,Dwyer:2014fpa,Evans:2021gyd}, and may be key for modeling 
intermediate-mass-ratio inspirals (IMRIs), $q^{-1}\sim 10^{-2}$--$10^{-4}$, a regime which remains challenging for NR~\cite{Healy:2017mvh,Fernando:2018mov,Lousto:2020tnb,Lousto:2022hoq}. 
It is especially exciting when considering recent detections of systems with $q^{-1}\sim 10^{-1}$, such at GW190814~\cite{LIGOScientific:2020zkf,LIGOScientific:2020ibl,LIGOScientific:2021usb} and GW191219~\cite{LIGOScientific:2021djp}.
These lie in a challenging regime where current models are not well-calibrated to NR simulations.

In this study we tackle an important limitation of previous analyses. 
In all cases, the agreement between NR and SMR predictions breaks down as the binary approaches the innermost stable circular orbit (ISCO).
This is expected: first because higher-order SMR coefficients may grow at high frequencies, but
also because previous SMR predictions expand around adiabatic inspiral, and must eventually fail as the binary transitions into plunge and merger.
Distinguishing between these two effects is crucial for modeling the binary near merger and is and particularly relevant for IMRIs and comparable-mass systems, such as those currently observed by ground-based GW detectors, where the transition region is large.

The transition can be understood as a singular boundary layer in between the slow inspiral through a sequence of circular orbits and a direct plunge with timescale $T \sim M$, $M$ the total mass.
In the SMR approximation, this transition occurs over a region of characteristic size $\sim M\nu^{2/5}$ around the location of the ISCO $r_* = 6M$, and with a dynamical timescale $T \sim M\nu^{-1/5}$.
Following the initial description of the transition dynamics~\cite{Ori:2000zn,Buonanno:2000ef}, a number of studies have refined and generalized the analytic approximations to the dynamics~\cite{Buonanno:2005xu,Sundararajan:2008bw,Kesden:2011ma,Taracchini:2014zpa,Apte:2019txp,Compere:2019cqe,Burke:2019yek}.
Recently GSF corrections have been incorporated into a generic expansion of transition equations and their solutions \cite{Compere:2021iwh,Compere:2021zfj}, which we use here.

Here we investigate the binding energy $E$ of nonspinning, quasicircular binary black holes at comparable masses and show that $E$ follows well-behaved SMR inspiral and transition expansions even through ISCO.
During the inspiral we recover the geodesic and post-geodesic coefficients as functions of in invariant radius $r_\Omega \equiv M^{1/3} \Omega^{-2/3}$. 
During the transition, we find that $E$ 
follows the expected fractional power expansion in $\nu$~\cite{Compere:2021zfj}, with the coefficients functions of the rescaled transition radius
$R_\Omega\equiv \nu^{-2/5}(r_\Omega-r_*)$.
This expansion breaks down near merger and towards inspiral as expected, and we estimate the region of validity to be $-2 \lesssim R_\Omega/M \lesssim 7$, corresponding to $r_\Omega \sim r_* -2 M \nu^{2/5}$.
The leading $\mathcal{O}(\nu^{4/5})$ coefficient from the transition fits is in good agreement with predictions.
We extract higher-order coefficients up to $\mathcal{O}(\nu^{9/5})$, where unknown 2GSF contributions first appear, and we find them to be negligible within the uncertainty of our analysis. 
Thus, our results indicate that an SMR expansion 
can provide accurate predictions for gravitational waves for comparable mass systems up to the final GW cycle before merger,
consistent with recent success of EMRI surrogate models~\cite{Rifat:2019ltp,Islam:2022laz} and 2GSF-accurate inspiral waveforms~\cite{Wardell:2021fyy}. 
Our analysis also shows that the transition constitutes a large portion of many signals observed by current ground-based detectors, indicating that an SMR approximation scheme augmented by transition dynamics, may have direct application to GW astronomy in the near future.

From here we set $G = c = M = 1$ and use
$'$ for derivatives with respect to $r_\Omega$ or $R_\Omega$, depending on the context. 
Quantities evaluated at ISCO are indicated by $*$. 

\prlsection{NR simulations} 
We select a set of high-resolution, nonspinning and quasicircular binary black hole simulations 
produced with the Spectral Einstein Code (SpEC)~\cite{Boyle:2019kee,SpECwebsite} 
with mass ratios ranging from $q=1$ to $q=20$. 
These simulations have low initial eccentricity $e \lesssim 10^{-4}$, a relatively large number of orbital cycles $N_{\rm cycles} \sim 20$--$45$, and in most cases two resolution levels, which allows us to assess numerical uncertainties. 

From each simulation we take the GW strain $h$ extrapolated to infinity~\cite{Boyle:2009vi,Boyle:2019kee} and corrected for the binary center of mass motion~\cite{Boyle:2015nqa,Boyle_scri_2020}.
From the strain we define the invariant radius $r_\Omega$ using an orbital frequency $\Omega$ inferred 
from the $\ell=2$, $m=2$ mode of the gravitational waves, $h_{22}$.
Although the quantity of interest for our analysis is $E$, only the energy flux $\dot E$ is directly accessible from the strain.
Thus, we analyze the gradients $E'(r_\Omega)=\dot{E}/\dot{r}_\Omega$ during inspiral and $E'(R_\Omega)=\dot{E}/\dot{R_\Omega}$ during the transition as an indirect measure of $E$.

We find that $h_{22}$ 
exhibits small modulations beyond those expected from quasicircular inspiral, which become particularly noticeable in $\dot \Omega$.
While the origin of these modulations is uncertain, during early inspiral they are dominated by residual junk radiation 
and at later times appear to be due to modulations of the center of mass, see e.g.~\cite{Albalat:2022lfz}.
To mitigate them, we apply a low-pass filter to $\dot \Omega$ during the early inspiral, with a cutoff frequency chosen conservatively high so that the overall chirping of $\dot \Omega$ is not biased.
Towards the transition regime the dynamics are fast enough that the filtering can still potentially bias the result. 
For $r < 9.5$ we smooth the modulations with a rolling fit of $E'(R_\Omega)$ to a quadratic over a fiducial window size of $\Delta R_\Omega = \pm 2$.
Further details of these procedures are in the Supplementary Materials.

\prlsection{Inspiral expansion}
During the adiabatic inspiral, post-geodesic corrections to $E$ can be calculated from 1GSF corrections to the redshift factor $z$, an invariant quantity constructed from the conservative piece of the metric perturbation \cite{Detweiler:2008ft}. 
The connection between $E$ and $z$ is a consequence of the first law of binary mechanics (FLBM) \cite{Friedman:2001pf,LeTiec:2011ab}, which assumes a helical symmetry with killing vector field $K=\partial_t+\Omega\partial_\phi$.
While this symmetry does not hold for dynamical binaries, the FLBM has been found to be surprisingly accurate when comparing analytic predictions to NR, e.g.~\cite{LeTiec:2011dp,Zimmerman:2016ajr}.
In our comparison we require the $O(\nu)$ corrections to the derivative of the binding energy, $E'(r_\Omega)$. 
For that, we use $E$ and $z$ expressions given in Ref.~\cite{LeTiec:2011dp} and translate these directly into predictions for
$E'(r_\Omega)$ 
as detailed in the Supplementary Materials.
Without assuming the relationships derived from the FLBM, 2GSF information is required to compute $\mathcal O(\nu)$ contributions to $E$~\cite{Pound:2019lzj}. 
We compare our NR result to both of these predictions.

\prlsection{Inspiral results}
\label{sec:resultsinspiral}

\begin{figure}[tb]
\includegraphics[width = 0.95\columnwidth]{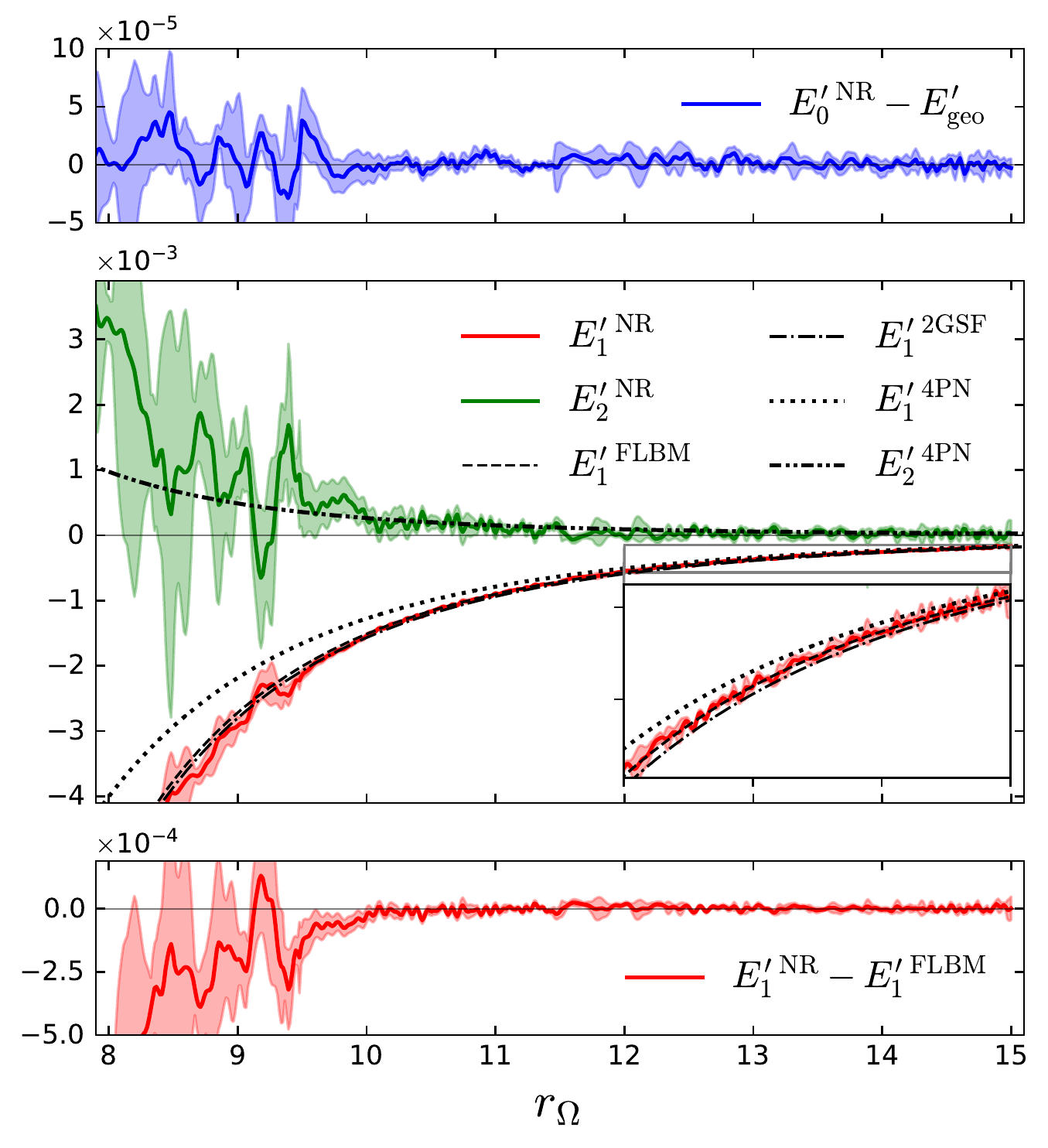}
 \caption{{\it Top panel}: Difference between 
 the leading coefficient $E_0'$ from a free fit to our simulations and the geodesic limit, showing the recovery of the energy gradient $E'(r_\Omega)$ from the data during inspiral.
 {\it Middle panel}: Inspiral results when using the geodesic limit as a baseline for our fit. We plot the subleading coefficients $E'_1$ (solid red) and $E'_2$ (solid green) of the inspiral expansion of $E'$. Also plotted is the corresponding $E'^{\rm{4PN}}_1$ prediction~\cite{LeTiec:2017ebm} and two SMR results: one based on the FLBM~\cite{LeTiec:2011dp} and one a post-adiabatic expansion including 2GSF corrections~\cite{Pound:2019lzj}. 
 {\it Bottom panel:} Difference between the NR and FLBM results for $E'_1$.}
 \label{fig:a1a2}
\end{figure}

To compare NR and SMR approximations during the inspiral, we perform a least-square fit of $E'(r_\Omega)$ to an expansion in integer powers of $\nu$,
\begin{align}
    E'(r_\Omega)=\sum_{i=0}
    E'_i(r_\Omega) \nu^{i},
    \label{eq:dEdr}
\end{align}
Following the approach of \cite{vandeMeent:2020xgc,Albalat:2022lfz}, we first fit the NR data at fixed $r$ values to both a first and second degree polynomial in $\nu$, without reference to the SMR prediction.
From this we extract values for the coefficients as a function of $r_\Omega$, and we recover the geodesic prediction $E'_0(r_\Omega)$  from the NR data alone.
This is shown in the top panel of Fig.~\ref{fig:a1a2}, where we plot the difference of $E'_0$ and $E_{\rm geo}$, finding remarkable agreement.

Having confirmed the test particle limit, we fit the remainder $\nu^{-1}[E'(r_\Omega)-E'_{\rm geo}(r_\Omega)]$ which allows us to extract the $\mathcal{O}(\nu)$ and $\mathcal{O}(\nu^2)$ coefficients more accurately.
Figure~\ref{fig:a1a2} shows the fitted coefficients $E'_1$ and $E'_2$.
The result is in good agreement with the first-order prediction from the FLBM, with systematic deviations starting $r_\Omega\lesssim 10$. 
The NR data agrees better with the FLBM than with the post-adiabatic result \cite{Pound:2019lzj} for $E$ based on a 2GSF calculation. 
This shows the importance of understanding subtle differences in the definitions of energy and orbital frequency that are used when comparing NR and GSF methods, see~\cite{Pound:2019lzj}. 
We also find evidence of a small but non-zero $\mathcal O(\nu^2)$ term during the inspiral.

We have repeated the inspiral analysis using the orbital angular momentum $L'$ derived from the angular momentum flux. 
As with the binding energy, a fit to $\mathcal O(\nu^2)$ provides excellent agreement throughout the inspiral regime, with a breakdown for $r_\Omega\lesssim 10$.
We also find that the adiabatic condition for circular orbits is satisfied for each of our extracted coefficients during inspiral, with $E'_i/L'_i = \Omega$ to within the uncertainty in our fits, including the $\mathcal O(\nu^2)$ coefficients.
This agreement provides further evidence for the accuracy of the FLBM during inspiral.

\prlsection{Transition expansion}

Using an SMR expansion around the Schwarzschild metric, the binding energy and radius of the orbit during the transition take the form~\cite{Compere:2021zfj}
\begin{align}
\label{eq:Etrans}
E&=E_*+\Omega_*[\nu^{4/5}\xi(\nu,s)+\nu^{6/5}Y(\nu,s)] \,,\\
\label{eqrEtrans}
r&=r_*+\nu^{2/5}R(\nu,s) \,.
\end{align}
where $s\equiv\nu^{1/5}(\tau-\tau_*)$ is the transition time parameter. 
The transition variables can be expanded in fractional powers of $\nu$: $\xi=\sum \xi_{i}\nu^{i/5}$,  $Y=\sum Y_{i}\nu^{i/5}$, $R=\sum R_{i}\nu^{i/5}$.
The transition equations provide a method for iteratively solving for each of $\xi_{i}$, $R_{i}$ and  $Y_{i}$, with their boundary conditions fixed by matching to inspiral at early times $s\rightarrow -\infty$. 
They also take as input the self-force $F_\mu$ in the neighborhood of $r_*$.
For example, angular momentum conservation reveals
$\xi=F^1_{\phi *} s + \mathcal{O}(\nu^{2/5})$
~\cite{Compere:2021zfj}.
For the self-force $F^1_{\phi*}$, we use first-order flux data at the ISCO, taken from~\cite{Taracchini:2014zpa,BHPToolkit}.
Note that our gauge-invariant $R_\Omega$ differs from $R$ at $\mathcal {O}(\nu^{2/5})$, and we have re-expanded the small parameter $q^{-1}$ in terms of $\nu$ which alters the usual transition expansion at $\mathcal{O}(\nu^{9/5})$,
where 2GSF corrections appear.  
This means that our final fitted transition functions differ from those of~\cite{Compere:2021iwh,Compere:2021zfj} beyond the leading order.
For our analysis we numerically solve for the leading order terms $R_0(s)$ and $\xi_0(s)$.
The leading transition equations we use are in the Supplementary Materials.

\prlsection{Transition results}
\label{sec:resultstransition}

\begin{figure}[tb]
\includegraphics[width = 0.95\columnwidth]{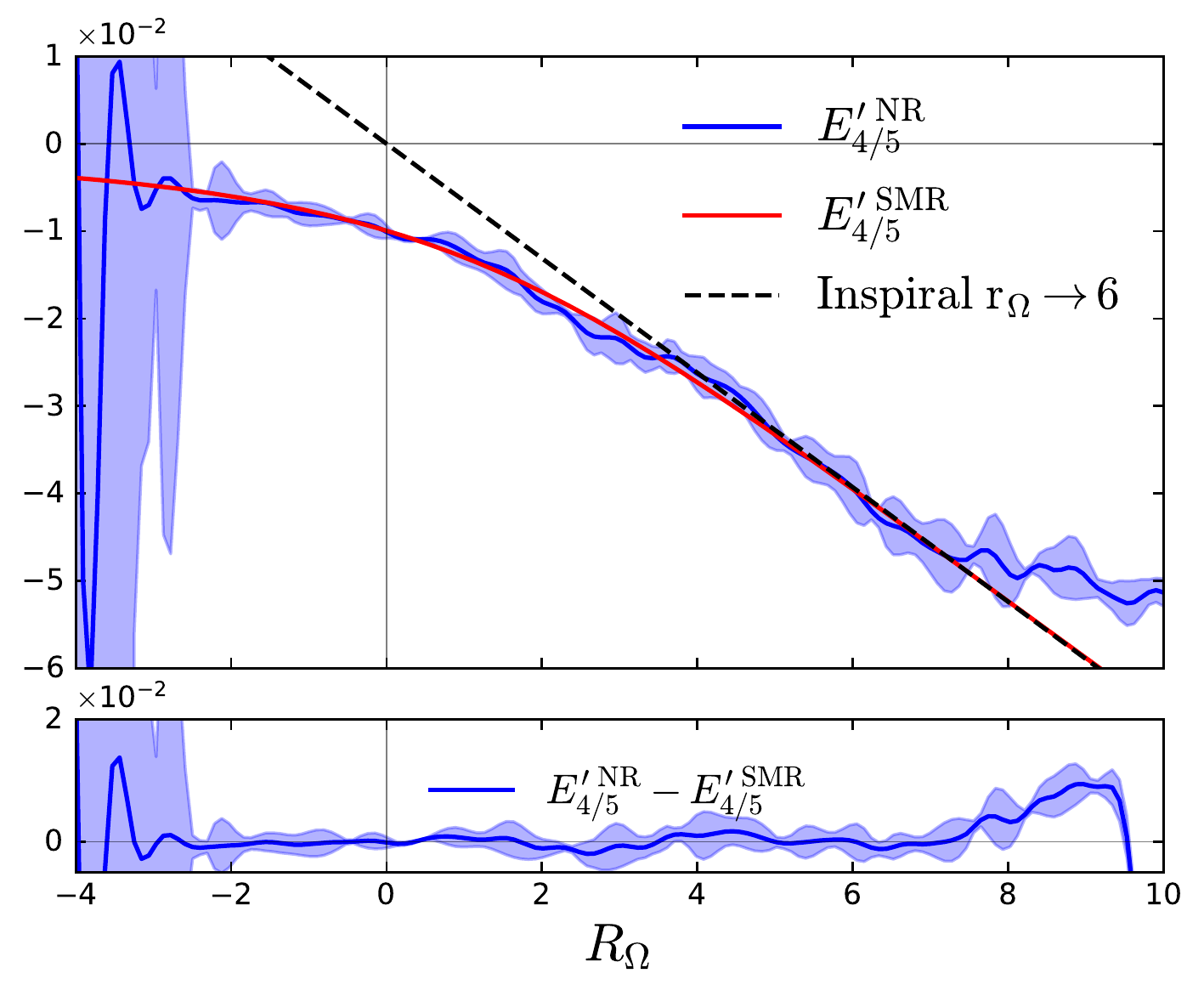}
 \caption{{\it Top panel}: Leading order coefficient $E'_{4/5}(R_\Omega)$ of the transition expansion of $E'(R_\Omega)$ obtained from the NR fit (blue) compared to the leading-order SMR prediction (red). 
 The recovery of the analytic prediction gives direct evidence for transition dynamics in comparable-mass NR simulations.
 {\it Bottom panel}: Difference between the SMR transition prediction and the NR result.}
 \label{fig:c4}
\end{figure}

\begin{figure}[tb]
\includegraphics[width = 0.95\columnwidth]{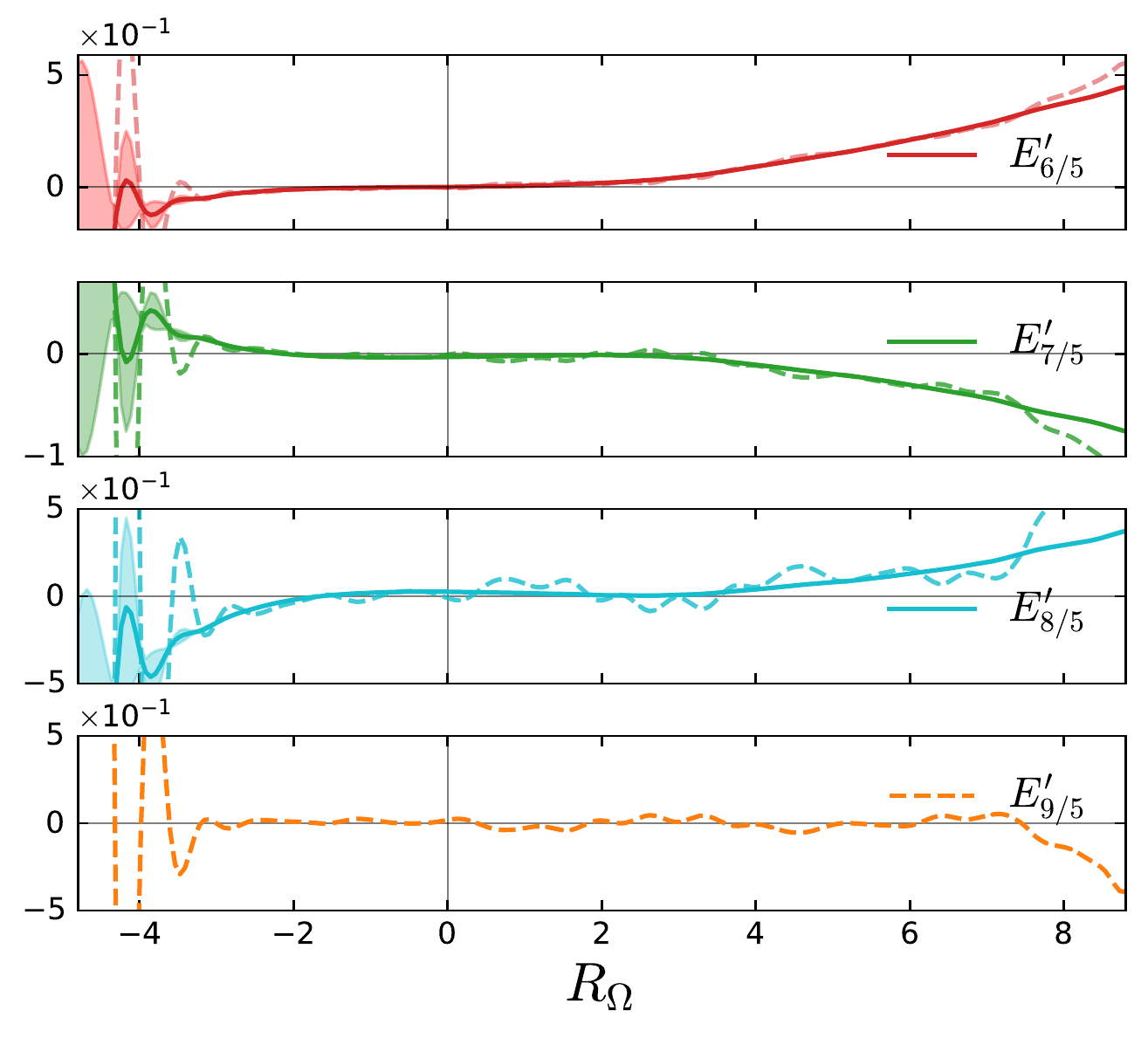}
 \caption{Subleading coefficients of the transition expansion for $E'(R_\Omega)$ from a fit to the NR data, fixing the first two coefficients.
 The solid lines correspond to the result of the fit with terms up to $\mathcal{O}(\nu^{8/5})$, and include our estimated uncertainties. 
 The dashed color lines represent the result of the fit including a $\mathcal{O}(\nu^{9/5})$ term where 2GSF effects first enter. 
 This term (bottom panel) is consistent with zero within our uncertainties.}
 \label{fig:c6c7c8c9}
\end{figure}

For the transition analysis we follow the same method as for the inspiral but we fit the NR data at fixed $R_\Omega$ to the fractional power expansion
\begin{align}
    E'(R_\Omega)&=\sum_{i=4}^{i_{\rm max}} 
    E'_{i/5}(R_\Omega) \nu^{i/5}.
    \label{eq:dEdR}
\end{align}
Following the expectation from Eq.~\eqref{eq:Etrans}, we set $E'_{5/5}=0$ in our first fit and let $i_{\rm{max}}=8$, 
which includes all transition orders where only 1GSF effects are present.
We find this number of terms is required to guarantee the stability of the fit and leave no structure in the residuals.
Figure~\ref{fig:c4} shows the results of the transition coefficients $E'_{4/5}$ from this fit.
The leading SMR result is in excellent agreement for 
$- 2 \lesssim R_{\Omega} \lesssim 7$, confirming the predicted transition dynamics even at comparable masses.

Having recovered the leading prediction $\nu^{4/5}\Omega_* d\xi_0/dR_0$ purely from the NR data, we next fit the residual between $E'$ and it.
In principle, the term we subtract is not accurate through $\mathcal {O}(\nu)$ and can introduce a term at $E'_{5/5}$.
As such, we first fit the NR data including this coefficient, and find the result is fully consistent with $E'_{5/5} = 0$
~\footnote{
This result is potentially surprising, since it seems to imply that the leading radial self-force effect, $f^r_{[0]}$ in \cite{Compere:2021iwh,Compere:2021zfj}, vanishes so that the $\mathcal{O}(\nu^{1/5})$ term in $R$, $R_1$, can be set to zero along with $Y_1$.
We speculate this may be because we work with the gauge-invariant $R_\Omega$ and the energy directly, while the piece $f^r_{[0]}$ sourcing $R_1$ is instead gauge-dependent.
}.
We then set $E'_{5/5} = 0$ and fit the scaled residual $[E'-\nu^{4/5}E^{\rm{SMR}}_{4/5}]\nu^{-2/5}$.
This allows us to extract accurate coefficients using either $i_{\rm max} = 8$ or $i_{\rm max} = 9$, with the latter providing an estimate for the $\mathcal{O}(\nu^{9/5})$ term.

Figure~\ref{fig:c6c7c8c9} shows the resulting higher-order coefficients. 
Generally the coefficients are comparable to $E_{4/5}(R_\Omega)$ in a region around ISCO but grow at larger $R_\Omega$, consistent with a breakdown of the transition expansion towards inspiral.
The coefficient $E'_{7/5}$ is consistently larger than $E'_{6/5}$ and similar to $E'_{8/5}$, which is why we require terms up to $E'_{8/5}$ to recover the leading-order result.
The leading result alone is never accurate at these mass ratios.
Including $E'_{9/5}$ results in clear overfitting of the residuals, and itself is consistent with zero, which further demonstrates that truncating the series at $i_{\rm max} = 8$ is appropriate when describing $E'$ using a transition expansion.
Our expansion fails for $R_{\Omega} \gtrsim 7$, as is clear from the failure to recover the leading-order result,
and from the blow-up of the subleading coefficients.
Similar results for the angular momentum and the variable $Y$ which describes the departure from circularity are given in the Supplementary Materials.
The latter analysis demonstrates that the combination of higher order transition terms $E'_{6/5} - \Omega_* L'_{6/5}$ are also in agreement with analytic predictions.

\prlsection{Conclusions}\label{sec:results}

\begin{figure}[tb]
\includegraphics[width = 0.98\columnwidth]{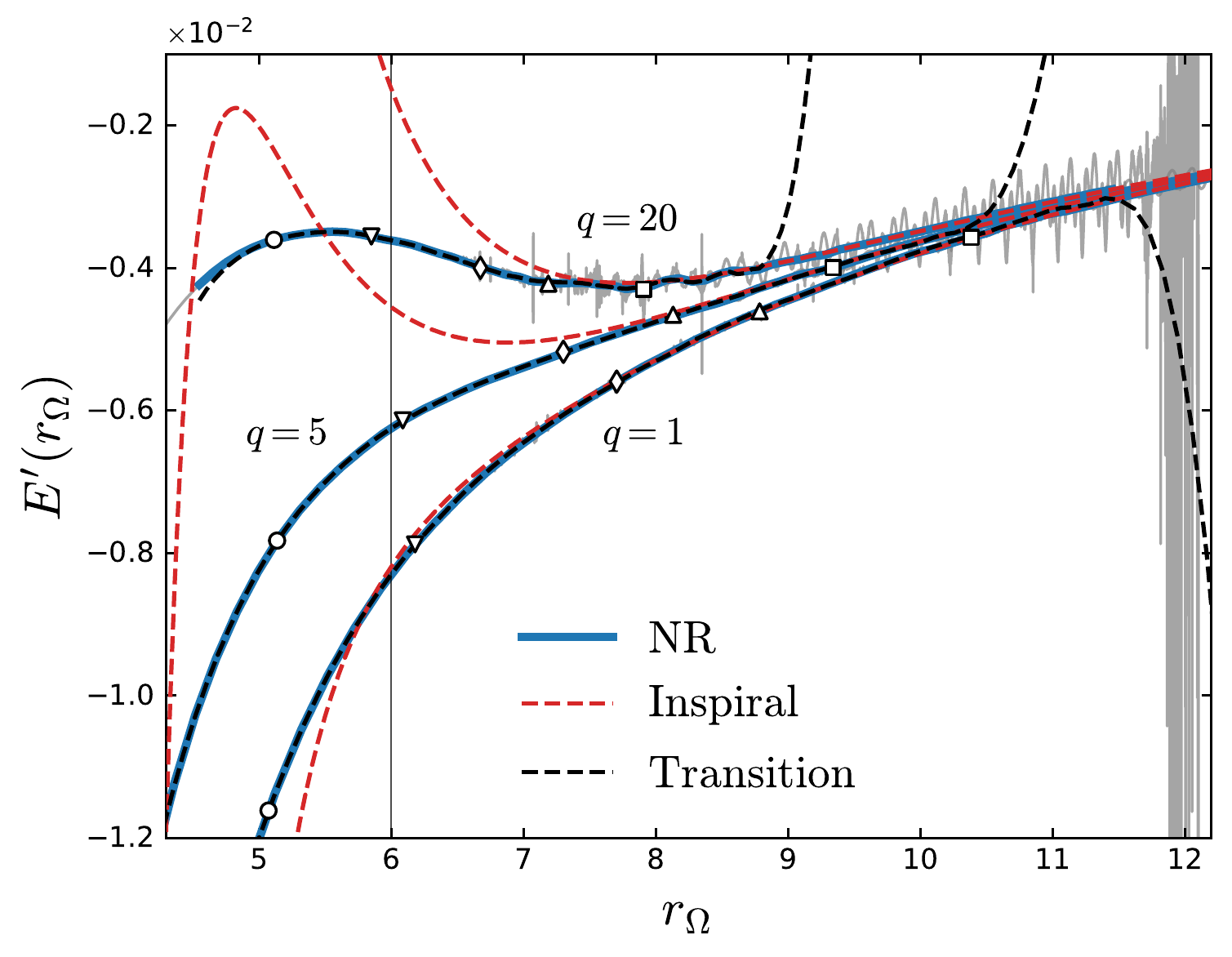}
 \caption{Depiction of the domains of applicability of each expansion for a sample of mass ratios $q=1,$ $5$, $20$. 
 We plot $dE/dr_\Omega$ from the NR data after removing the oscillations (blue), the raw NR data (light gray), the inspiral prediction from the FLBM (dashed red lines), and the transition expansion resulting from our fit, excluding $\mathcal{O}(\nu^{9/5})$ and higher) (black dashed lines). The markers indicate the number of GW cycles left before merger: 1 (circle), 2 (down-triangle), 4 (diamond), 6 (up-triangle) and 10 (square).
 A combination of the analytic inspiral approximation and transition dynamics including 1GSF information models the binding energy accurately up until the final $1$-$2$ GW cycles.
 }
 \label{fig:matching}
\end{figure}

We have extracted for the first time the SMR limit from nonspinning, quasicircular NR simulations in the transition region around the ISCO.
Our work extends previous analyses of the validity of the SMR approximation at comparable masses, which were restricted to the inspiral region (but see also~\cite{Islam:2022laz}).
We find that an adiabatic SMR expansion, together with the FLBM, is in good agreement with our simulations for $r_\Omega \gtrsim 10$.
The failure of the FLBM result at smaller radii can be explained by the onset of transition dynamics.
Using a transition expansion for the binding energy and angular momentum as functions of $R_{\Omega}\equiv \nu^{-2/5}(r_{\Omega}-r_{\rm{isco}})$, we can 
recover the leading-order SMR result~\cite{Ori:2000zn,Buonanno:2000ef} in a region of width $-2 \lesssim R_{\Omega} \lesssim 7$ around the ISCO. 
We find that terms up to $\mathcal{O}(\nu^{8/5})$ are necessary to recover this result. We also give a prediction for the value of the higher-order coefficients and show that the $\mathcal{O}(\nu^{9/5})$ contribution is zero to within our uncertainties, suggesting a small 2GSF contribution to $E$.

Our results are summarized in Fig.~\ref{fig:matching}, which shows the NR data for $dE/dr_\Omega$, for three binaries with $q=1$, $q=5$, and $q=20$.
We compare our raw NR data with the smoothed and filtered data that we fit, along with the $\mathcal{O}(\nu)$-accurate FLBM inspiral prediction and the results of our transition fit up to $\mathcal{O}(\nu^{8/5})$.
This illustrates the failure of the inspiral treatment near ISCO for higher $q$, the narrowing of the transition region with increasing $q$, and the fact that a combination of the two treatments describes the energy accurately until the last cycle before merger in all cases using only 1GSF information. 

The next step would be to explore transition contributions to the GW phasing, extending the results of~\cite{vandeMeent:2020xgc} to merger.
SMR predictions in this regime would be enabled by combining transition modeling~\cite{Compere:2021iwh} with 2GSF-accurate fluxes~\cite{Warburton:2021kwk} and waveforms~\cite{Wardell:2021fyy}.
It then is critical to include spins 
using an SMR expansion around Kerr.
Another direction would be to examine eccentric binaries.
These areas represent the frontier of 2GSF calculations.
If achieved, GSF could provide a complete, first-principles model for the two-body problem, applicable from EMRIs to equal masses.

\prlsection{Acknowledgements}
We would like to thank the authors of the SpEC simulations used in this analysis: Serguei Ossokine, Joohean Yoo, Vijay Varma and Jonathan Blackman. 
We also thank Adam Pound, Niels Warburton, Barry Wardell and Leanne Durkan for discussions about the transition expansion and filtering method and for generously sharing the 2GSF flux data from~\cite{Warburton:2021kwk}.
For the simulations used in this work, computations were performed on the Wheeler cluster at Caltech, which is supported by the Sherman Fairchild Foundation and by Caltech; and on Frontera at the Texas Advanced Computing Center~\cite{frontera}.
We also thank the developers of \texttt{Scri} \cite{Boyle:2015nqa,Boyle_scri_2020}, which was used to calculate the energy and angular momentum fluxes.
This work makes use of the Black Hole Perturbation Toolkit~\cite{BHPToolkit}.
S.N.A.~and A.Z.~are supported by NSF Grant Numbers PHY-1912578 and PHY-2207594. 
M.G.~is supported by NSF Grant Number PHY-1912081 at Cornell.
M.A.S.~is supported in part by the Sherman Fairchild Foundation and by National Science Foundation (NSF) Grant Nos. PHY-2011961, PHY-2011968, and OAC-1931266 at Caltech.

\appendix

\section{Supplementary materials}
\label{sec:AppxDetails}

\prlsection{Details of numerical simulations} 
Table~\ref{table:1} gives further details on the numerical simulations used in this study, which includes $q=14$ and $q=15$ simulations associated with the surrogate described in~\cite{Yoo:2022erv}.
The initial data types correspond to Superposed Kerr Schild (SKS)~\cite{Lovelace:2008tw} and Superposed Harmonic Kerr (SHK)~\cite{Varma:2018sqd}.
The initial data makes use of improved prescriptions to minimize the initial center of mass motion~\cite{Ossokine:2015yla}, and the initial eccentricity is reduced using an iterative method~\cite{Pfeiffer:2007yz,Buonanno:2010yk,
Mroue:2012kv}. 
The resolution level is indicated by the Lev argument, and corresponds to internal tolerances of the adaptive mesh refinement algorithm used in SpEC~\cite{Lovelace:2010ne,Szilagyi:2014fna}.
Due to the nature of adaptive mesh refinement, and the variable history of refinement for each simulation, strict spectral convergence is not expected for derived quantities such as the gravitational wave strain $h$~\cite{Boyle:2019kee}.
For this reason, we assess numerical uncertainties by including simulations with different resolutions where available in our error estimates.
From each simulation we take the extrapolated gravitational wave strain $h= h_+ - i h_\times$ at future null infinity~\cite{Boyle:2009vi,Boyle:2019kee}, expanded in $(\ell,m)$ modes of spin-weighted spherical harmonics.
We use an $N=4$ extrapolation setting (fifth-order in $r^{-1}$) as appropriate for inspiral~\cite{Boyle:2009vi,Boyle:2019kee} for all our simulations.
The strain $h$ is further corrected by applying a translation and boost that minimizes the effect of the center of mass motion present in the simulations~\cite{Boyle:2015nqa,Boyle_scri_2020}. 

The orbital frequency is estimated using the $(2,2)$ mode of the gravitational wave strain,
\begin{align}
\phi_{22} & = \arg h_{22} \,, & 
\Omega& =\dot \phi_{22}/ 2\,,
\end{align}
and the flux is computed using the standard formula
\begin{align}
\dot{E}&=\lim_{r\rightarrow \infty}\frac{r^2}{16\pi}\sum^{\ell,m} |\dot{h}_{\ell m}|^2,
\end{align}
summing over all modes with available to us from our simulations, $\ell \leq 8$.
In principle one can integrate the flux to find the binding energy, 
choosing the integration constant by matching either to the mass of the final black hole or to PN theory early in the inspiral. 
We find that this procedure introduces undesired errors in the analysis, which is why we focus on the gradient of the binding energy in our analysis.

\prlsection{Details of the data analysis and error estimates}
During inspiral, we use a second-order forward and backward Butterworth filter $\mathcal B$ to filter $\dot \Omega$.
To mitigate the impact on the frequency sweep, we find that find that applying the filter after subtracting a good estimate of the data improves its performance by reducing the overall variation of the data.
For this we use the 2GSF prediction for $\dot \Omega$ from~\cite{Wardell:2021fyy}.
Further, we apply the filter to $\dot \Omega$ as a function of its index.
Since the time step in our simulation is adaptive, the strain data $h$ is not uniformly sampled, but we find that resampling to uniform time steps makes finding an appropriate cutoff frequency more challenging than for the non-uniform sampling.
The reason is that with the denser sampling rate at late times effectively brings the late-time chirp to lower frequencies, so that a uniform cutoff frequency better targets the actual noisy behavior at all times.
The cutoff frequencies are chosen as $f_c = a + b (\nu-\nu_{q=20})$, with $a\in \{6,8,10\}\times10^{-4}$ and $b = 1.95\times10^{-3}$.
The filtered data is then
\begin{align}
    \dot{\Omega}_{{\rm filtered},i}=[\mathcal B * (\dot{\Omega}_{{\rm raw},i}-\dot{\Omega}_{{\rm SMR},i})] + \dot{\Omega}_{{\rm SMR},i} \,. 
\end{align}

Since the filtering process biases the data late in the simulation, and because we have no SMR data beyond ISCO, we choose a cutoff radius $r_{\Omega,c}=9.5$ beyond which we simply switch to a rolling fit of $E'(\Omega)$ (or $L'(R_\Omega)$ as discussed below) to a quadratic, over a window $\Delta R_{\Omega} \in \pm\{1,2,3\}$ around the fitted point.
For our fiducial analysis, we select the one corresponding to $\Delta R=2$ and $a=8 \times 10 ^{-4}$.

To create our uncertainty bands for our fitted quantities, we vary all of the parameters involved in the filtering and smoothing of our data within the stated ranges, as well as repeating our analysis with simulations at a lower resolution. 
The error bands are created by taking the envelope of the variation in our fitted parameters, over the different resolutions, and the $1$-$\sigma$ errors of our least-squares fits at each frequency point.

\begin{table}[t]
\centering
\setlength{\tabcolsep}{0.6em}
\begin{tabular}{ccccccccc}
\toprule
    $q$ & Type & $M\Omega_0$ & $N_\text{cycles}$ & $e_0$ & $\text{Levs}$ & SXS ID
    \\\midrule
    1 & SKS &0.01233 & 27.96 & 1.355e-4 & $\text{5,6}$ & 2513  
    \\
    1.5  & SKS &0.01250 & 28.98 & 5.77e-5  & $\text{2,3}$ & 2331
    \\
    2  & SHK &0.01554 & 20.70 & 2.408e-4 & $\text{2,3}$ & 2497 
    \\
    2.5 & SKS & 0.01512 & 22.49 & 7.580e-4 & 2,3 & 0191 
    \\
    3  & SHK &0.01707 & 20.44 & 9.64e-5 & $\text{2,3}$ & 2498 
    \\
    3.5  & SKS &0.01477 & 27.76 & 2.665e-4 & $\text{4,5}$ & 2483
    \\
    4  &SKS  &0.01600 & 25.67 & 8.702e-4 & $\text{4,5}$  & 2485 
    \\
    4.5  & SKS &0.01616 & 27.37 & 8.289e-4 & $\text{4,5}$ & 2484 
    \\
    5  & SKS &0.01589 & 29.13 & 2.236e-4 & $\text{4,5}$ & 2487  
    \\
    5.5  & SKS &0.01592 & 30.81 & 4.442e-4 & $\text{4,5}$ & 2486  
    \\
    6  & SKS &0.01588 & 32.62 & 5.864e-4 & $\text{4,5}$ & 2489  
    \\
    6.5  & SKS &0.01599 & 34.43 & 7.263e-4 & $\text{4,5}$ & 2488  
    \\
    7  & SKS &0.01577 & 36.16 & 3.612e-4 & $\text{4,5}$ & 2491 
    \\
    7.5  & SKS &0.01597 & 37.89 & 3.694e-05  & $\text{4,5}$ & 2490
    \\
    8  & SKS &0.01584 & 39.53 & 6.688e-4 & $\text{5}$ & 2493 
    \\
    8.5  & SKS &0.01594 & 41.31 & 8.578e-4 & $\text{5}$ & 2492 
    \\
    9  & SKS &0.01583 & 43.16 & 2.010e-4 & $\text{4,5}$ & 2495 
    \\
    9.5  & SKS &0.01585 & 44.93 & 1.584e-4 & $\text{4}$ & 2494 
    \\
    14 & SHK & 0.02292 & 27.70 & 3.814e-4 & $\text{2,3}$ & 2480 
    \\
    15 & SHK & 0.02317 & 27.94 & 3.692e-4 & $\text{2,3}$ & 2477 
    \\
    20 & SKS &  0.02321 & 34.38 & 2.506e-4 & $\text{3,4}$ & 2516 
    \\
    \bottomrule
\end{tabular}
\caption{Details of the nonspinning quasicircular SpEC simulations used in this analysis. The subscript zero denotes the reference time (time at which junk radiation has sufficiently decayed).}
\label{table:1}
\end{table}

\prlsection{Details of the inspiral prediction}
The FLBM gives~\cite{LeTiec:2011ab,LeTiec:2011dp} 
\begin{align}
    E'(r_\Omega)&=E'_{\rm{geo}}(r_\Omega)+\nu E'_{\rm{FLBM}}(r_\Omega)+\mathcal{O}(\nu^2) \,,
\end{align}
with
\begin{align}
    E_{\rm{FLBM}}(x)&=\frac{1}{2}z_{\rm{1GSF}}(x)-\frac{x}{3}z'_{\rm{1GSF}}(x)-1 \notag \\
    &+\sqrt{1-3x}+\frac{x}{6}\frac{7-24x}{(1-3x)^{3/2}} \,,
\end{align}
and where $x\equiv r_{\Omega}^{-1}$.
The 1GSF contribution to $z$, $z_{\rm{1GSF}}(x)$, has been computed to high precision with multiple codes, e.g.~\cite{Dolan:2014pja}. For the purpose of our inspiral comparison we make use of the fit formula~\cite{LeTiec:2011ab} 
\begin{align}
\label{eq:z1}
    z_{\rm{1GSF}}(x)=\frac{2x(1-2.18522 x + 1.05185 x^2)}{1 - 2.43395 x + 0.400665 x^2 - 5.9991 x^3}.
\end{align}

\prlsection{Details of the transition formalism}
Our analysis requires that we solve the transition equations to leading order, in order to calibrate our fits and compare with the NR results.
The quantities $R_{0}(s)$ and $Y_{0}(s)$ are solved using the leading order transition equations \cite{Ori:2000zn,Buonanno:2000ef,Kesden:2011ma,Compere:2021zfj}
\begin{align}
    \left(\frac{dR_{0}}{ds}\right)^2&=-\frac{2}{3}\alpha_* R^3_{0}-2\beta_*\kappa_*s R_{0}+\gamma_* Y_{0},\\
    \frac{d^2R_{0}}{ds^2}&=-\alpha_* R^2_{0}-\kappa_* \beta_* s,\\
    \frac{dY_{0}}{ds}&=2\kappa_*\frac{\beta_*}{\gamma_*}R_{0}.
\end{align}
The constants in the above equations are given by
\begin{align}
    \kappa_* &\equiv F_{\phi}^1\rvert_{r_*}\,, 
    \qquad  \alpha_* \equiv \frac{1}{4}\frac{\partial^3 V^{\rm{geo}}}{\partial r^3}\biggr\rvert_{\rm{isco}} \,, 
    \qquad \gamma_* \equiv \frac{\partial V^{\rm{geo}}}{\partial L}\biggr \rvert_{r_*}\,, \\
    \beta_* &\equiv -\frac{1}{2} \left( \frac{\partial^2 V^{\rm{geo}}}{\partial r\partial L} + \Omega\frac{\partial^2 V^{\rm{geo}}}{\partial r \partial E}\right)\biggr\rvert_{r_*}\,,
\end{align}
where $V^{\rm{geo}}$ is the effective potential of radial geodesic motion about a Schwarzschild black hole.

\prlsection{Further results: Angular momentum}

Here we present the analysis of the angular momentum in both the inspiral and transition regimes.
These results provide an independent demonstration of the accuracy of the SMR modeling during inspiral and plunge. 
They are used to confirm that the binaries are described by adiabatically evolving circular orbits through $\mathcal O(\nu^2)$ during inspiral, consistent with the FLBM, as discussed in the main text.
We also find that this condition fails at the expected order during transition, by finding agreement with the non-circular correction $Y_0$ introduced by Kesden~\cite{Kesden:2011ma}.

We extract the angular momentum flux from our numerical simulations using the standard formula
\begin{align}
    \dot{L}&=\lim_{r\rightarrow \infty}\frac{r^2}{16\pi}\text{Im}\left(\sum_{l,m} m h^{l,m} (\dot{h}^{l,m})^* \right) \,.
\end{align}
We use the same filtering technique applied to $\dot \Omega$ as done for the analysis of $E(r_\Omega)$ during the inspiral, and the same rolling fit to $dL/dR_\Omega$ as a function of the transition radius $R$ for $r_\Omega < 9.5$, as done for analysis of the binding energy.

\begin{figure}[tb]
\includegraphics[width = 0.95\columnwidth]{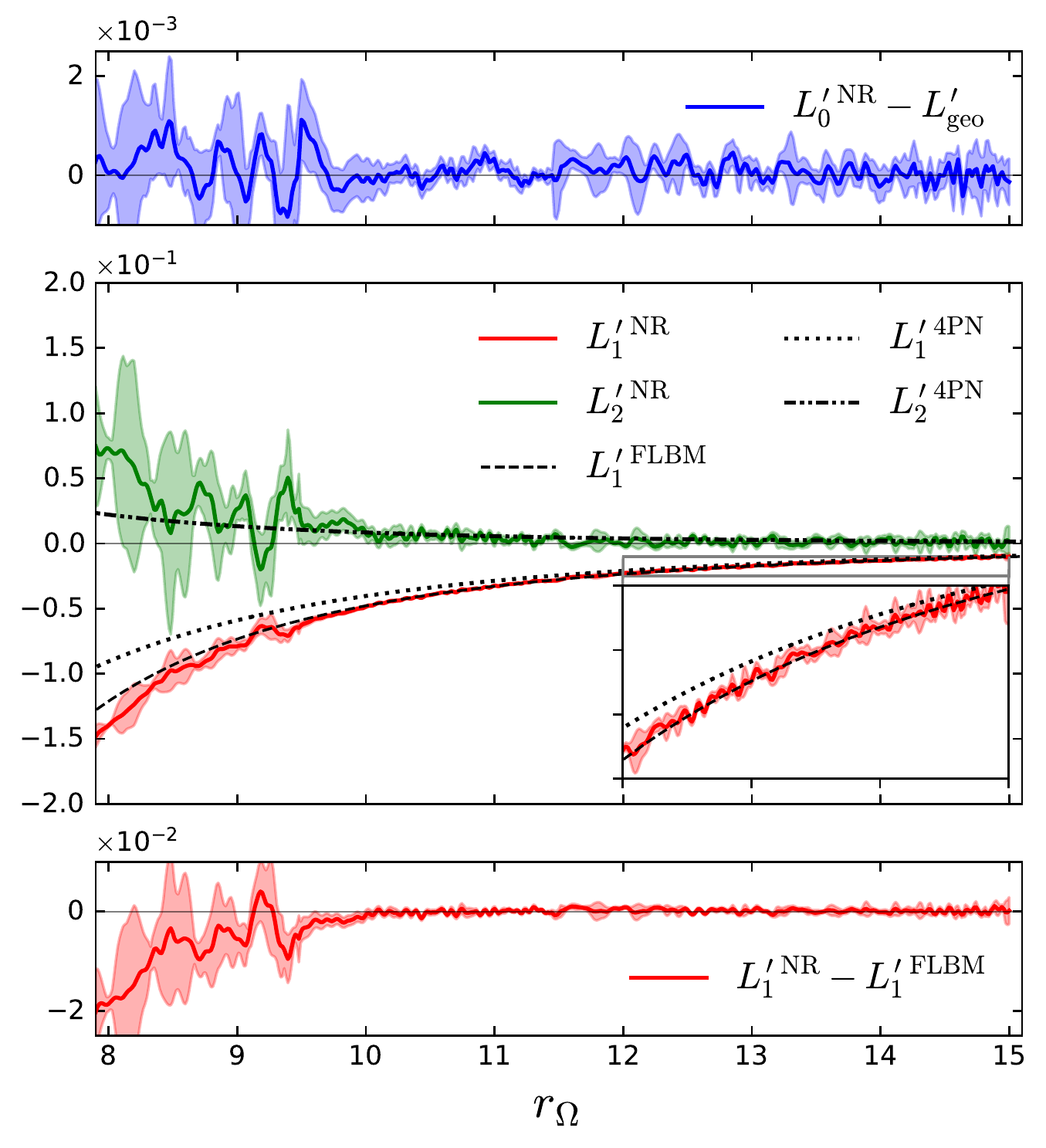}
 \caption{{\it Top panel}: Recovery of the geodesic limit of the energy gradient $L'(r_\Omega)$ during inspiral from the NR data. We plot the difference between the leading coefficient $L_0'$ from a free fit to our sequence of NR simulations and the geodesic limit, together with our estimated uncertainties.
 {\it Middle panel}: Inspiral results when using the geodesic limit as a baseline for our fit. We plot the subleading coefficients $L'_1$ (solid red) and $L'_2$ (solid green) of the inspiral expansion of $L'$. Also plotted is the corresponding $L'^{\rm{4PN}}_1$ prediction~\cite{LeTiec:2017ebm} and SMR result based on the FLBM~\cite{LeTiec:2011dp}.
 {\it Bottom panel:} Difference between the NR and FLBM results for $L'_1$.}
 \label{fig:b1b2}
\end{figure}

For the analytic comparisons during inspiral, we use the FLBM-derived expansion relations~\cite{LeTiec:2011ab,LeTiec:2011dp}
\begin{align}
L'(r_\Omega)&=L'_{\rm{geo}}(r_\Omega)+\nu L'_{\rm{FLBM}}(r_\Omega)+\mathcal{O}(\nu^2)\,,\\
L_{\rm{FLBM}}(x)&=\frac{1}{3\sqrt{x}}z'_{\rm{1GSF}}(x)+\frac{1}{6\sqrt{x}}\frac{4-15x}{(1-3x)^{3/2}} \,,
\end{align}
and the redshift factor of Eq.~\eqref{eq:z1}.
During inspiral we use an expansion in integer powers of $\nu$ to fit to the NR data,
\begin{align}
 L'(r_\Omega)=\sum_{i=0}^{2} L'_i(r_\Omega) \nu^{i}.
 \label{eq:dLdr}
\end{align}
As before, the fits across simulations at fixed $r_\Omega$ give a leading coefficient $L'_0$ in agreement with the geodesic prediction $L'_{\rm geo}$, and so we subtract this and fit the residual $L'(r_\Omega) - L'_0(r_\Omega)$ to improve the accuracy of the fitted $L'_1(r_\Omega)$ and $L'_2(r_\Omega)$ coefficients.
The results of the inspiral analysis are depicted in Fig.~\ref{fig:b1b2}.

\begin{figure}[tb]
\includegraphics[width = 0.95\columnwidth]{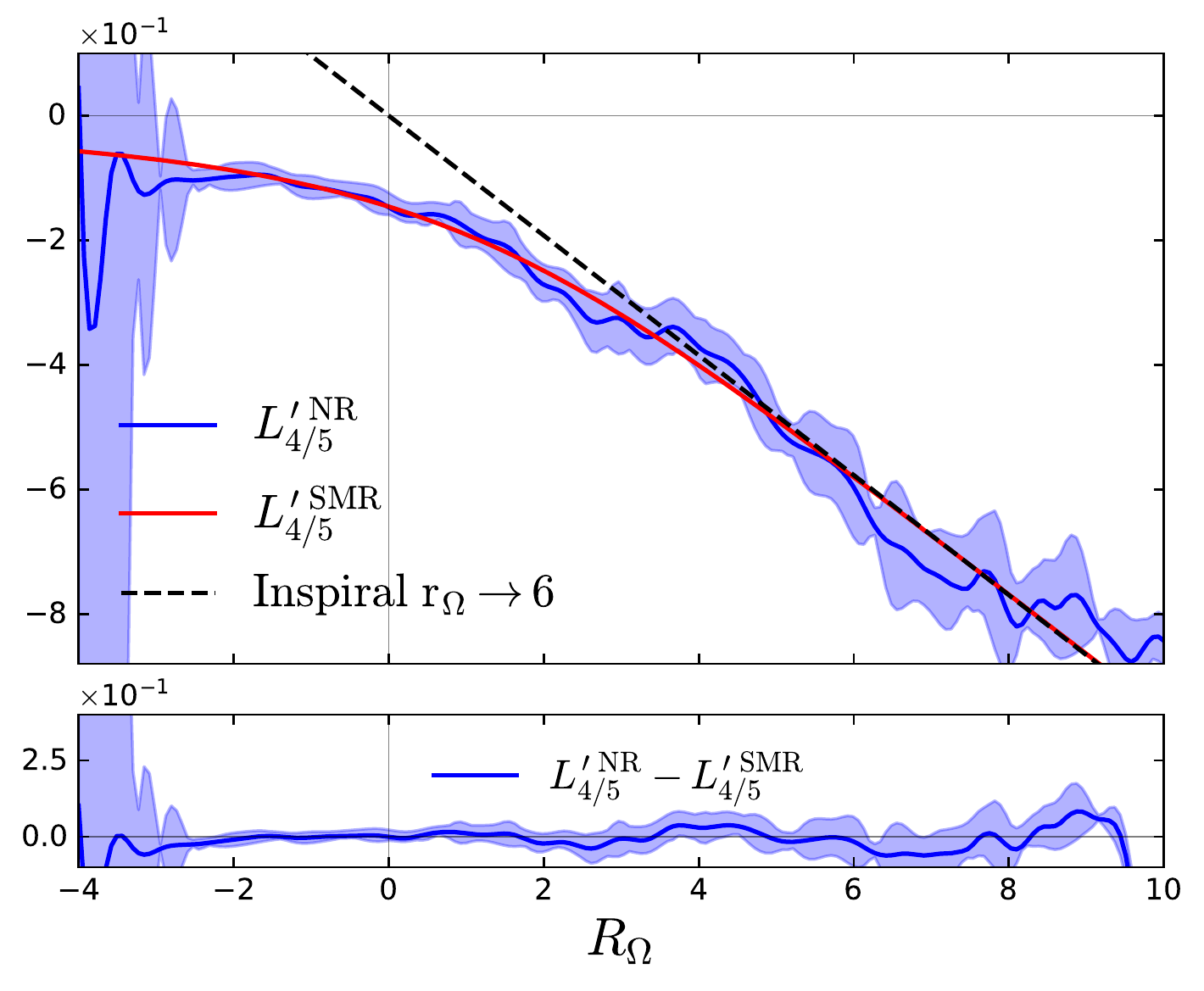}
 \caption{{\it Top panel}: Leading order coefficient $L'_{4/5}(R_\Omega)$ of the transition expansion of the binding energy obtained from the NR fit (blue) compared to the leading-order SMR prediction (red). {\it Bottom panel}: Difference between the SMR transition prediction and the NR result.}
 \label{fig:d4}
\end{figure}

\begin{figure}[tb]
\includegraphics[width = 0.95\columnwidth]{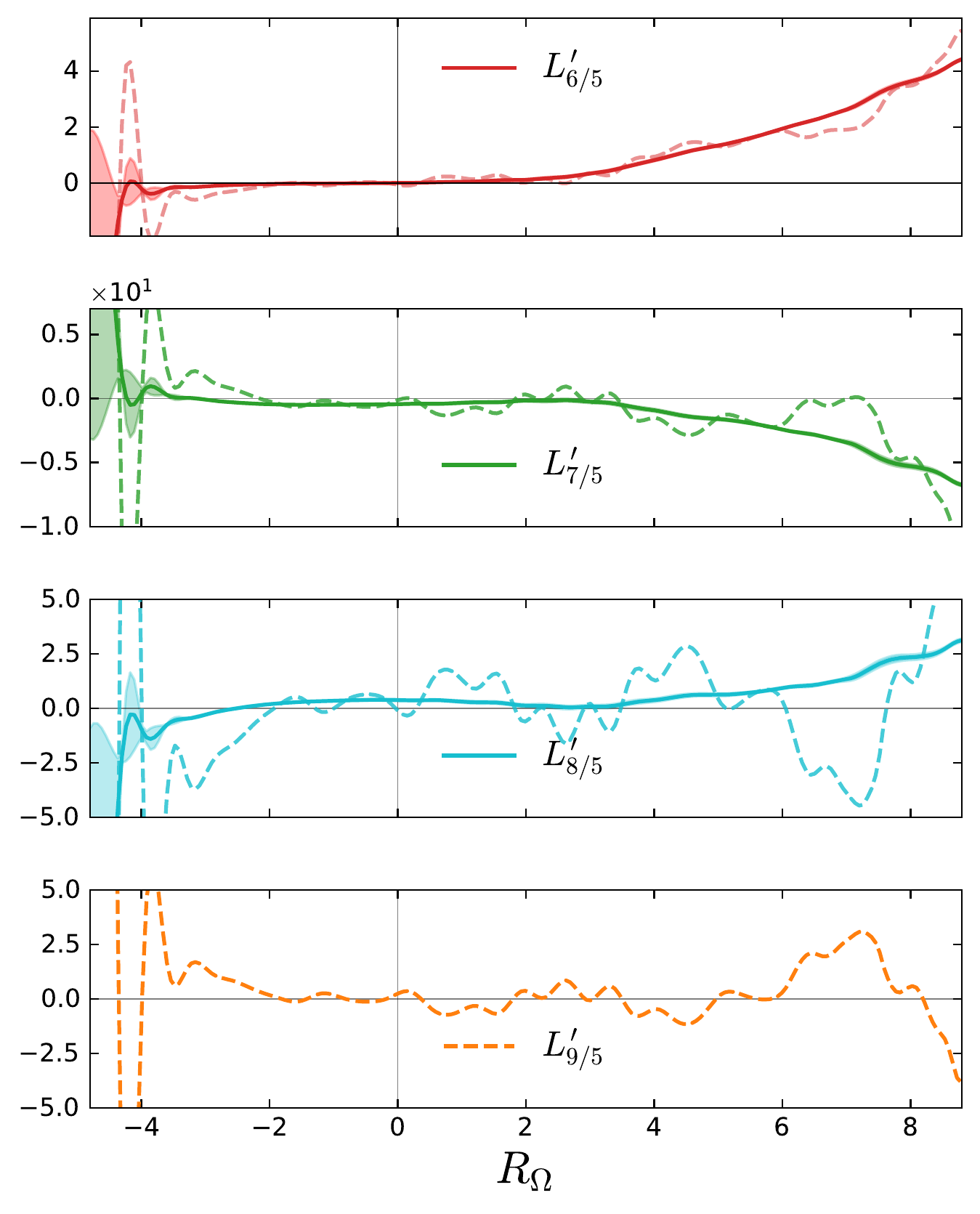}
 \caption{Subleading coefficients of the transition expansion of $L'(R)$ the NR analysis. The solid lines correspond to the result of the fit with terms up to $\mathcal{O}(\nu^{8/5})$. The dashed color lines represent the result of the fit including a $\mathcal{O}(\nu^{9/5})$ term shown in the bottom panel.}
 \label{fig:d6d7d8d9}
\end{figure}

For the transition analysis, the analytic approximation for $L'(R)$ depends on the quantity $\xi$~\cite{Ori:2000zn,Compere:2021iwh,Compere:2021zfj},
\begin{align}
        L&=L_*+\nu^{4/5}\xi(\nu,s)\label{eq:Ltrans} \,.  
\end{align}
This allows us to compare $L'(R_\Omega)$ during the transition to analytic predictions in the same way as for $E'(R_\Omega)$.
We carry out the same analysis as for $E'(R_\Omega)$, first performing a fit of the form
\begin{align}
     L'(R_\Omega) &=\sum_{i=4}^{i_{\rm max}} L'_{i/5}(R_\Omega) \nu^{i/5},
\end{align}
with $i_{\rm max} = 8$ and the only from analytic theory being that $\xi_1'(R_\Omega) = 0$.
The results for the leading coefficient shown in Fig.~\ref{fig:d4} confirm that $L_{4/5}'$ obeys the expected transition dynamics, in agreement with theory.
Next we fix $L_{4/5}'$ to the analytic prediction, and verify that $L'_{5/5}$ remains consistent with zero.
Finally fixing both $L'_{4/5}$ and $L'_{5/5}$, we fit with $i_{\rm {max}} = 8$ and $i_{\rm max} = 9$ and extract the transition coefficients plotted in Fig.~\ref{fig:d6d7d8d9}.
As with our analysis of the binding energy, we see that the higher terms in the expansion are significant and well behaved through $-2 \leq R_\Omega \leq 7$, and that the $\mathcal{O}(\nu^{9/5})$ term is consistent with zero to within our uncertainties.
This again confirms the small size of 2GSF contributions to the transition.

\begin{figure}[tb]
\includegraphics[width = 0.98\columnwidth]{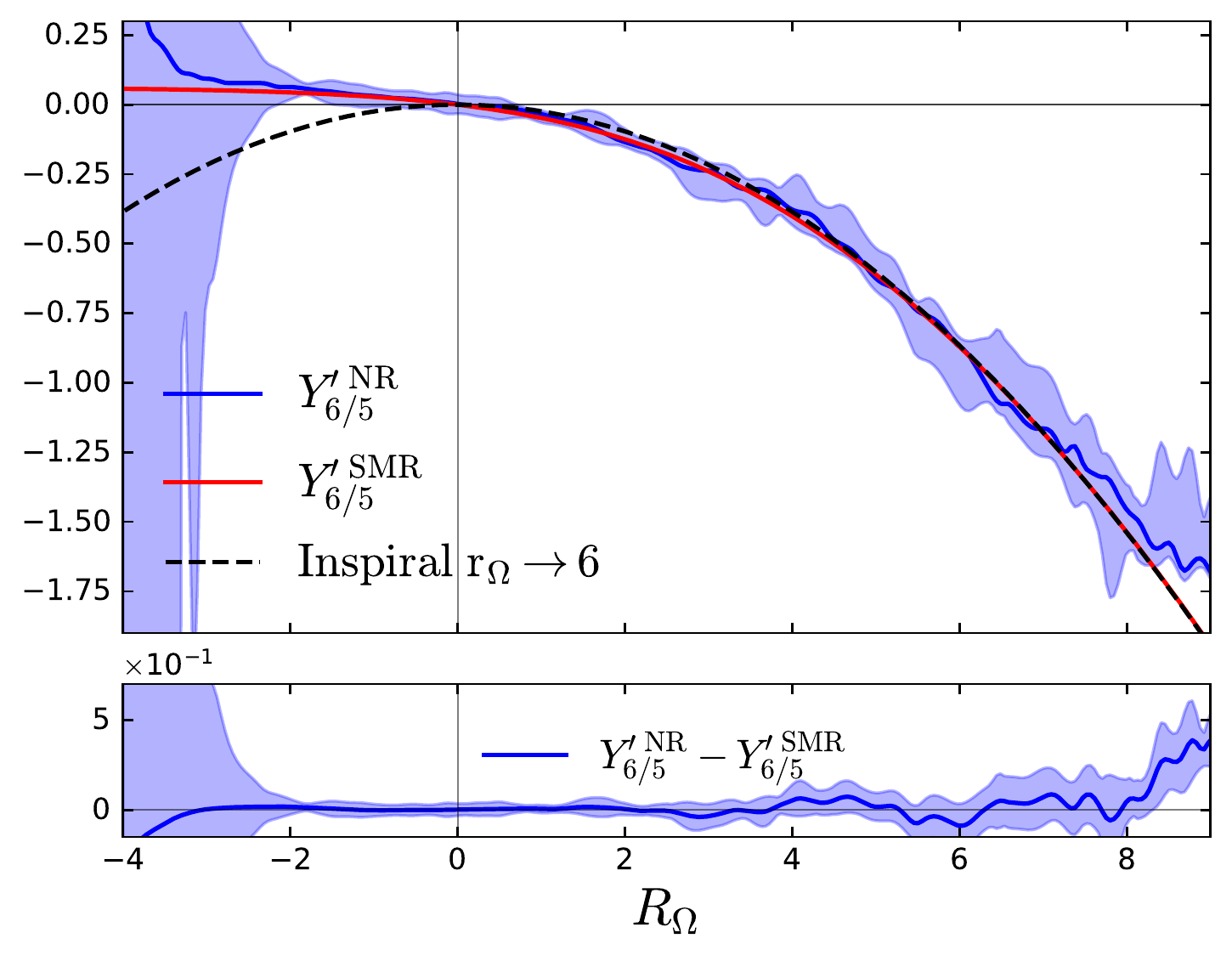}
 \caption{{\it Top panel}: Leading order coefficient $Y'_{6/5}(R_{\Omega})$ of the transition expansion obtained from the NR fit (blue) compared to the Kesden SMR prediction (red). {\it Bottom panel}: Difference between the SMR transition prediction and the NR result.}
 \label{fig:gamma}
\end{figure}

Equivalently, we can focus on the deviation from circularity during transition~\cite{Kesden:2011ma}, $Y'(R_\Omega) \equiv E'(R_\Omega) - \Omega_* L'(R_\Omega)$.
The leading prediction for the deviation is
\begin{align}
\label{eq:LeadingY}
    Y'(R) & = Y'_0(R_0)\nu^{6/5} + \mathcal{O}(\nu^{7/5}).
\end{align}
We fit the NR data during transition to a fractional power series of the form
\begin{align}
     Y'(R_\Omega) &=\sum_{i=6}^{i_{\rm max}} Y'_{i/5}(R_\Omega) \nu^{i/5}\label{eq:GammaExpand},
\end{align}
once again setting the first correction to the leading-order behavior to zero, $Y'_{7/5} = 0$, and letting $i_{\rm max} = 8$.
Figure~\ref{fig:gamma} shows the fit coefficient $Y'_{6/5}(R_\Omega)$ extracted from this procedure.
The leading-order prediction provides a good fit to the data over the same range of $R_{\Omega}$ values as we find for the energy analysis, $2 \lesssim R_{\Omega} \lesssim 7$.

If we retain the $Y'_{7/5}$ term, the near-degeneracy of the first two terms prevents us from achieving this level of agreement with theory for $Y'_{6/5}$.
We can justify setting $Y'_{7/5} = 0$ in two ways.
Our first comes from considering what our numerical results for $E'(R_\Omega)$ imply for the transition expansion.
The fact that we find that $E'_{5/5} = 0$ after subtracting the leading transition prediction from the NR data for $E'(R_\Omega)$ implies that the first correction to $R$ vanishes, $R_1(R_\Omega) = 0$.
This in turn implies that first term in the transition expansion of the radial self-force, $f^r_{[0]}$ of~\cite{Compere:2021zfj}, vanishes.
Without this term to source $Y_1$ in the transition equations, $Y_1$ term can be set to zero in a consistent manner.
The second way is to subtract the leading-order prediction~\eqref{eq:LeadingY} from $Y'$ and fit the remainder, $[Y'(R_\Omega) - \nu^{6/5} dY_0/dR_0]$ to the factional power expansion in $\nu$, starting the series with $Y'_{7/5}$.
When doing so we find the fit for $Y'_{7/5}(R_\Omega)$ vanishes throughout the transition region, while the higher coefficients are nonzero, similar to what occured for $E'(R_\Omega)$.

\bibliography{main.bbl}

\end{document}